\documentclass[conference]{IEEEtran}

\usepackage{amsmath}
\usepackage{amssymb}
\usepackage{graphicx}

\newtheorem{cor}{Corollary}
\newtheorem{defn}{Definition}
\newtheorem{problem}{Problem}
\newtheorem{thm}{Theorem}
\newtheorem{lem}{Lemma}

\begin{document}

\title{Delay-Energy lower bound on Two-Way Relay Wireless Network Coding}

\author{
\IEEEauthorblockN{Hongyi Zeng}\footnote{Hongyi Zeng was a student at Tsinghua University when this work was done.}
\IEEEauthorblockA{Department of Electrical Engineering\\
Stanford University\\
Stanford, CA 94305\\
Email: hyzeng@stanford.edu}
\and
\IEEEauthorblockN{Wei Chen}
\IEEEauthorblockA{Department of Electrical Engineering\\
Tsinghua University\\
Beijing, China 100084\\
Email: weichen@tsinghua.edu.cn}
}

\maketitle
\begin{abstract}
Network coding is a novel solution that significantly improve the
throughput and energy consumed of wireless networks by mixing traffic
flows through algebraic operations. In conventional network coding
scheme, a packet has to wait for packets from other sources to be
coded before transmitting. The wait-and-code scheme will naturally
result in packet loss rate in a finite buffer. We will propose Enhanced
Network Coding (ENC), an extension to ONC \cite{Chen_ICC07} in continuous
time domain. 

In ENC, the relay transmits both coded and uncoded packets to reduce
delay. In exchange, more energy is consumed in transmitting uncoded
packets. ENC is a practical algorithm to achieve minimal average delay
and zero packet-loss rate under given energy constraint. The system
model for ENC on a general renewal process queuing is presented. In
particular, we will show that there exists a fundamental trade-off
between average delay and energy. We will also present the analytical
result of lower bound for this trade-off curve, which can be achieved
by ENC.\end{abstract}

\markboth{Tsinghua University}{and this is for right pages}

\section{Introduction}

Network coding, where packets from two or more sources
are allowed to be transmitted and processed jointly, has recently
attracted much attention due to its potential for high-speed network
\cite{Ahlswede}. It has recently been found that the broadcast nature
of network coding is suitable for improving the energy consumed in
wireless network, which is critical in energy-limited wireless sensor
network. Significant recent effort has been dedicated to this field. 

Using network coding in wireless scenario for saving energy consumption
and improving information exchange consumed is first proposed in \cite{Wu}.
In \cite{Chen_ICC06}, a wireless network with a centralized relay
and $N$ sources exchanging their information, was proposed in order
to characterize the relay-assistant network under fading and multiuser
interference. A more general scenario without centralized relay is
presented in \cite{Chen_Globecom}. With network coding, the number
of packets needed transmitted is decreased significantly, which improves
energy consumed significantly. A physical layer network coding was
considered in \cite{Popovski,Zhang}. The received signal is simply
amplified and broadcast in a noise version of the summation of the
two source signals. By doing this, the complexity of the relay node
is reduced.

The above works were mainly focused on enhancing the performance of
a synchronized network. However, due to the stochastic nature of wireless
network, the packets arriving pattern should be taken into account.
Moreover, delay and packet-loss rate, which is critical parameter
in Quality of Service (QoS), is seldom considered. In \cite{Chen_ICC07},
delay-energy relation is considered, and theoretical delay-energy
trade-off curve is given. Furthermore, A novel scheduling scheme for
network coding, ONC - Opportunistic Network Coding, is proposed in
the discrete time domain in order to achieve the optimal trade-off
boundary. An analysis on first-come-first-serve policy in continuous
time domain is presented in \cite{He}. In this paper, ONC will be
extended to continuous time domain, referred as Enhanced Network Coding
(ENC), which mainly focuses on renewal process and Poisson process
for packet arriving pattern. We will show that with a finite buffer,
conventional network coding scheme results in inevitable packet loss.
This paper will also presented the optimal delay-energy trade-off
curve in continuous time domain. We will observe that first-come-first-serve
policy is not sufficient to achieve the minimized delay.

The rest of this paper is organized as follows. The details of system
model and general description of ENC policy is presented in 
\autoref{sec:System-Model}. \autoref{sec:Energy-Delay-Trade-off-Analysis}
will show that why we need ENC and the rigorous mathematical model
of ENC for renewal process and Poisson process arriving pattern. The
expression of systems parameters including delay, packet-loss rate,
and energy consumed is proposed. With the parameters given, 
\autoref{sec:Optimal-Delay-Power-Trade-off} relates delay and energy
with a linear programming problem. After solving this problem, we
will have the optimal curve of delay-energy trade-off as well as the
optimal ENC policy under different energy constraints. It will be
shown that there exists a fundamental trade-off between delay and
energy, which implies the performance limits of wireless network coding.

\section{System Model}
\label{sec:System-Model}

\begin{figure}
\centering
\includegraphics[scale=0.3]{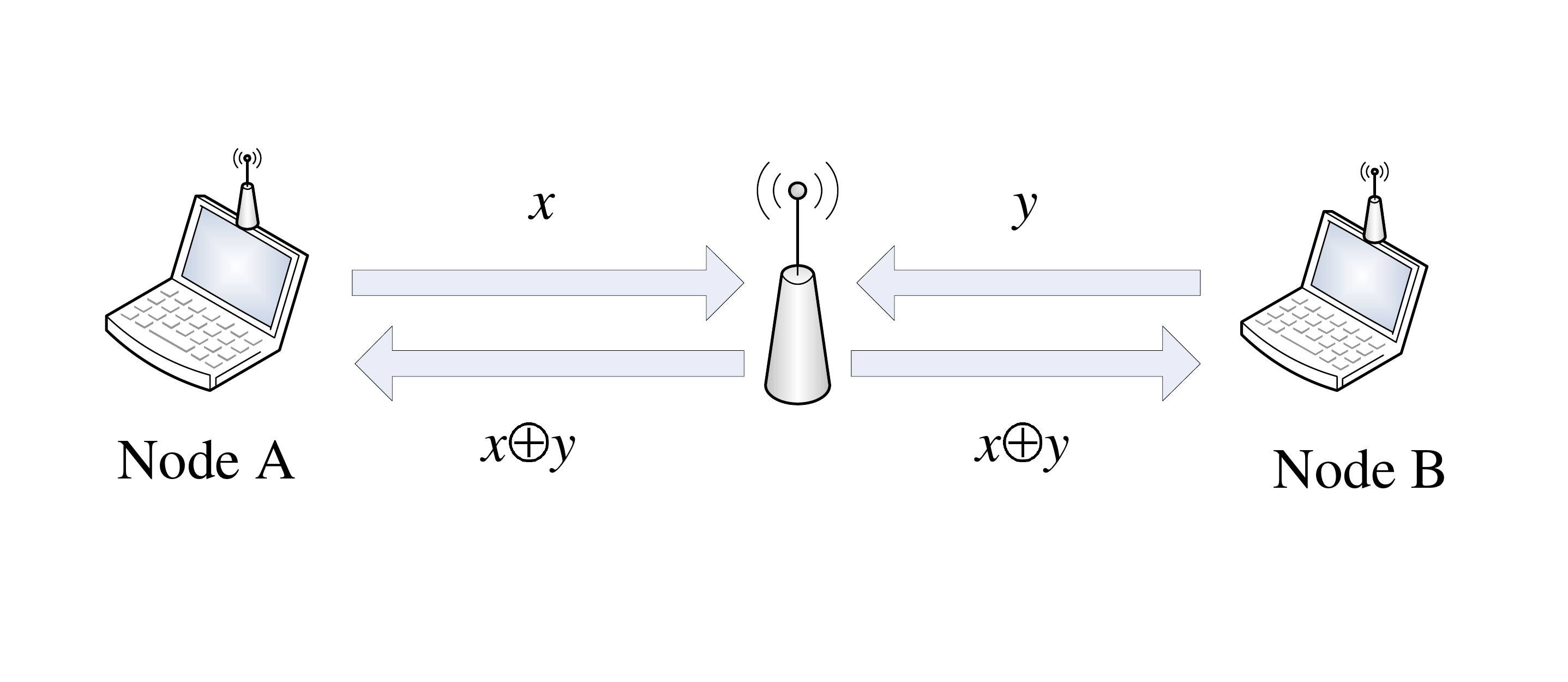}
\caption{Wireless Switching Network using Network Coding}
\label{fig:Wireless-Switching-Network}
\end{figure}

A wireless network consists two nodes A and B that wants to exchange
their packets via a relay is considered, as depicted in \autoref{fig:Wireless-Switching-Network}.
With network coding, the relay only need to broadcast the bit-by-bit
XOR result of a packet of Node A and a packet of Node B. Each node
can then decode its desired packet by implementing XOR operation again
between the received packet and the packet from itself. Compared with
the conventional method of transmitting the two packets individually,
network coding can save 50\% energy for the relay node in each transmission.

\begin{figure}[tb]
\centering
\includegraphics[scale=0.4]{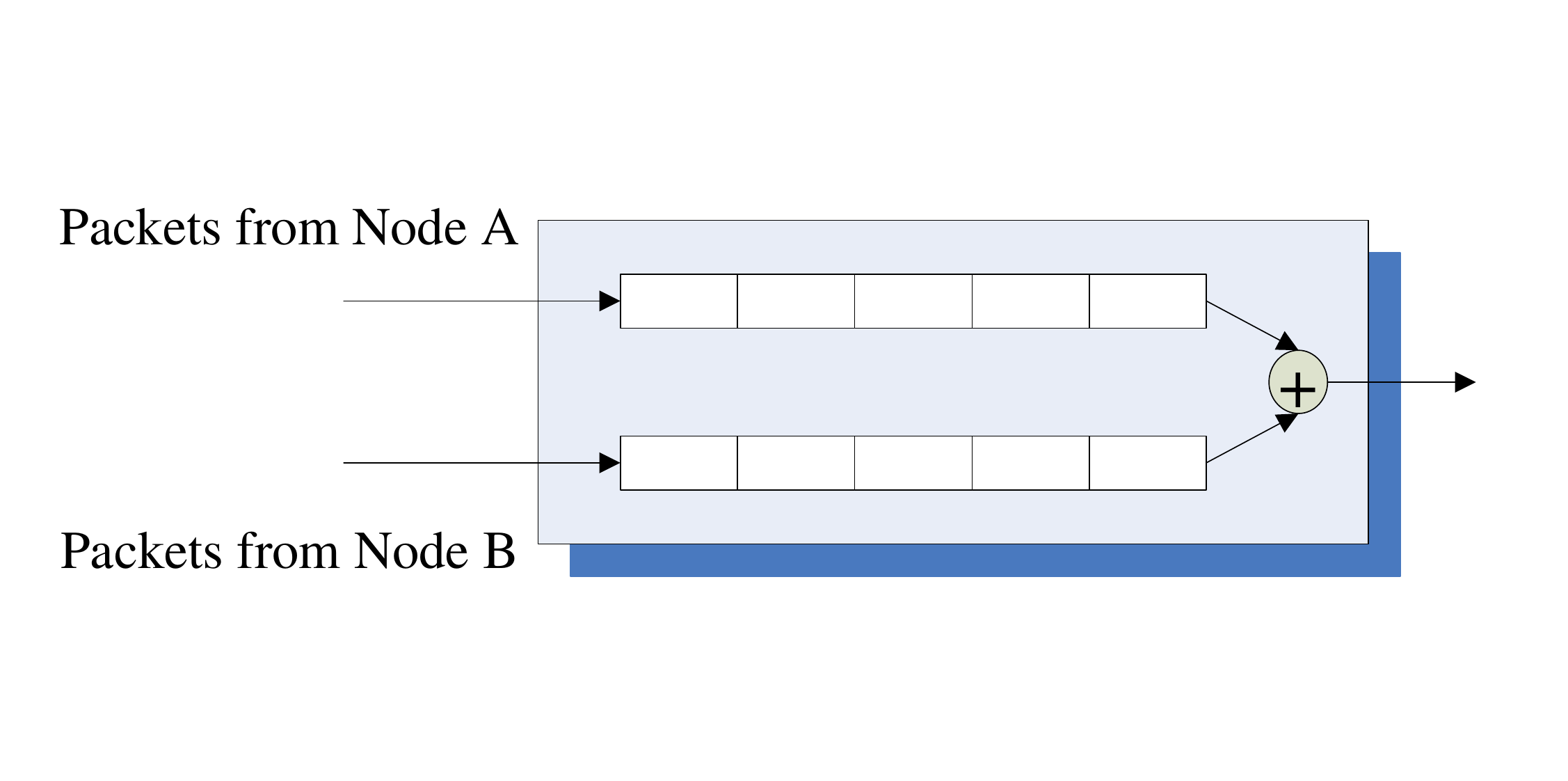}
\caption{Queuing Model in the Relay Node}
\label{fig:Queuing-Model-in}
\end{figure}

Packets from node A and B arrive at the relay from two independent
Poisson processes with equal parameters $\lambda$ through an ideal
channel. We will show that the parameter $\lambda$ could be also
applied in renewal process. The relay node maintains two finite buffers
to store the backlog packets, which can hold at most $K$ packets,
as shown in \autoref{fig:Queuing-Model-in}. The relay employs
the following policy in transferring the packet.

\begin{figure}
\includegraphics[scale=0.6]{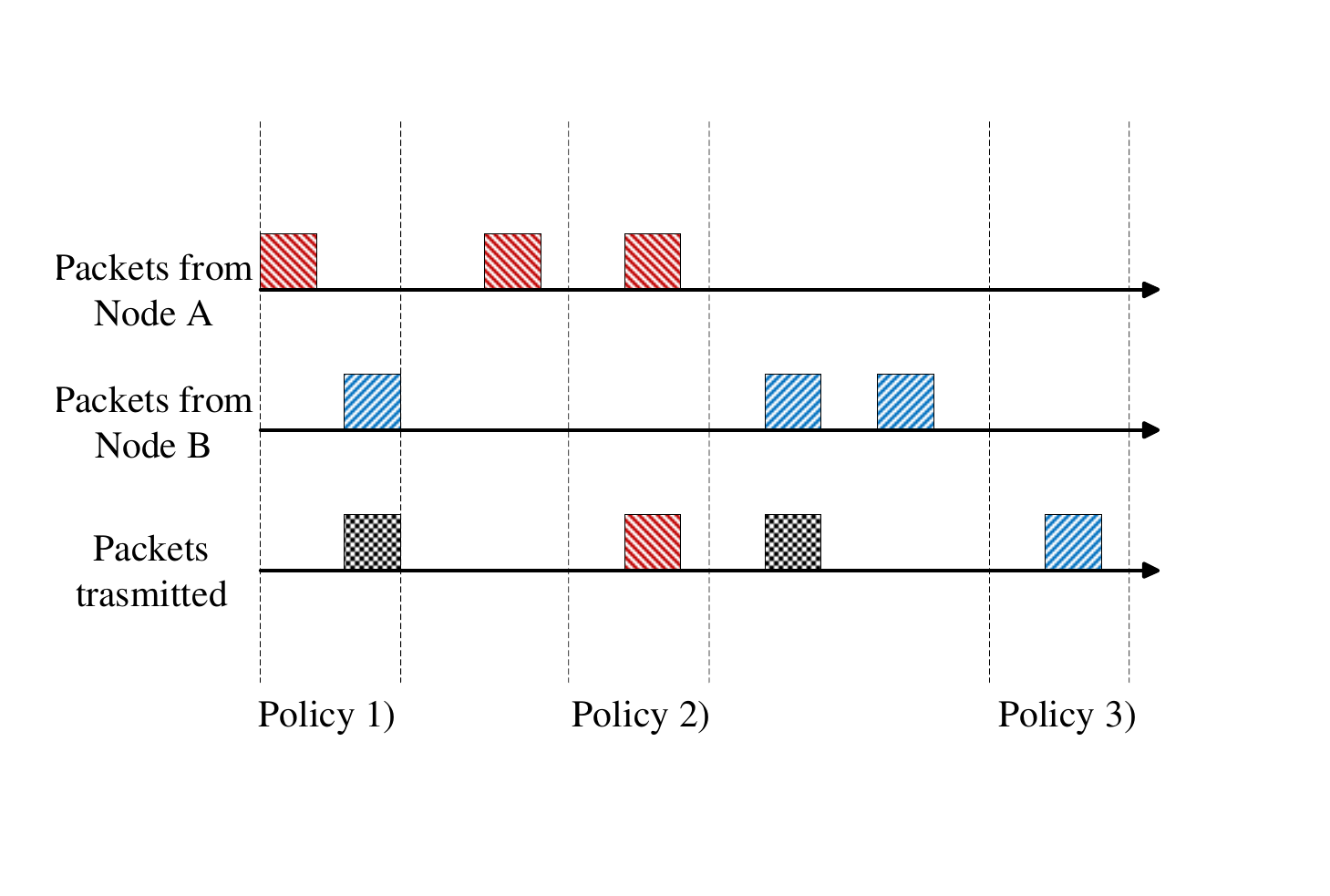}
\caption{ENC Policies}
\label{fig:ENC-Policies}
\end{figure}

\begin{enumerate}
\item When a packet arrives, if the queue holding the traffic from the opposite
direction is not empty, the relay sends out the coded version of the
binary sum of this packet and the packet in the queue from the opposite
direction \emph{immediately}.
\item When a packet arrives, if the queue holding the traffic from the opposite
direction is empty, the relay sends out the oldest packet in the queue
with the probability of $g_{k}$, where $k$ denotes the current number
of storing packet in the queue.
\item When no packet arrives, the relay sends out the packet it stored with
a certain probability. This probability is described as a Poisson
process depending on the current state $k$. The parameter for this
Poisson process is $f_{k}$.
\end{enumerate}
All of above policies are depicted in \autoref{fig:ENC-Policies}.
In all cases, we assume that the packet size is small enough that
the transmission delay is negligible. In other words, the transmissions
can be modeled as points on the time axis. Therefore, the average
delay experienced by the packets equals the average amount of time
the spend in the queue at the relay. In the next section, we will
explore the relationship of the average delay with the average transmission
energy of the relay.

Note that when the probability of sending an un-coded packet is fixed
to be 0, ENC reduces to conventional wireless network coding, where
all packets are network-coded. Hence, conventional wireless network
coding can be viewed as a special case of ENC. In \cite{He}, a policy
called ``first-come-first-serve''(FCFS) was presented. It can be
viewed as another special case of ENC with parameters $f_{k}\equiv0$.
The relay only transmits un-coded packet when buffer is full. There
is a brief summary in \autoref{tab:Policy-comparation-among}.

\begin{table}[tb]
\centering
\begin{tabular}{ccc}
\hline
ENC Parameters & $g_{k}$ & $f_{k}$\\
\hline
\hline 
Conventional Network Coding & 0 & 0\\
\hline 
``first-come-first-serve'' & $g_{k}=\begin{cases}
1 & k=K\\
0 & k\neq K
\end{cases}$ & 0\\\hline
\end{tabular}
\caption{Policy comparison among conventional network coding, FCFS and ENC}
\label{tab:Policy-comparation-among}
\end{table}

In this paper, we will show that conventional wireless network coding
is not optimal, in terms of delay and packet loss, even results in
huge packet-loss rate. The performance limit of wireless network coding,
in terms of delay-energy trade-off, will be presented. We shall also
optimize ENC to achieve the optimal trade-off.

\section{Delay-energy Trade-off Analysis}
\label{sec:Energy-Delay-Trade-off-Analysis}

In this section, we present a rigorous mathematical description of
ENC. A finite-state Markov chain is formulated for delay and energy
consumed analysis.

\subsection{Why ENC?}
\label{sub:Why-do-we}

Conventional network coding approach usually assumes a buffer with
a infinite length on the relay for storing packets. However, in the
practical implementation, infinite buffer is impossible. Through the
following theorem, we could see that the finite buffer implies the
conventional network coding has inevitable packet loss.

First of all, we remark that at most one queue at the relay node can
be non-empty. (Otherwise, the relay should XOR and transmit the packets
from two queues \emph{immediately}. We denote the number of packets
on the relay at the time instance $t$ by $|R(t)|$, and the number
of all packets arriving at the relay before the time instance $t$
by $Q(t)$. If $R(t)>0$, it means that only packets from Node A is
in the queue. If $R(t)<0$, packets from Node B can be found in the
queue. $B_{k}$ is a random variable represents the $k^{th}$ packets
that arriving at the relay. Its definition is as follows:
\begin{equation*}
B_{k}:=\begin{cases}
1 & \text{Packet comes from Node A}\\
-1 & \text{Packet comes from Node B}
\end{cases}
\end{equation*}

We assume that each $B_{k}$ has independent identical distribution
for all $k$ and $E(B_{k}=0$. 
\begin{thm}
The probability of buffer overflow is
\begin{equation*}
\Pr(|R(t)|>K)\approx2\Phi(-\frac{K}{\sqrt{Q(t)}\sigma_{B}}
\end{equation*}

for any given finite $K$-packet-buffer, where $\sigma_{B}$ is the
variance of $B_{k}$, and $\Phi(z)=\Pr(X<z)$, $X\sim N(0,1)$.\end{thm}
\begin{IEEEproof}
With the definition in \autoref{sec:System-Model}, since the
sum of $B_{k}$ represents the number of remaining packets in the
relay, one can easily obtains the following equation, 
\begin{equation*}
R(t)=\underset{k=1}{\overset{Q(t)}{\sum}}B_{k}
\end{equation*}
When $t\rightarrow\infty$, $Q(t)\rightarrow\infty$, according to
central limit theorem, we have
\begin{equation*}
R(t)\sim N(0,\sqrt{Q(t)}\sigma_{B}
\end{equation*}

Now let us calculate the probability of buffer overflow.
\begin{equation*}
\begin{array}{ccc}
\Pr(|R(t)|>K) & = & \Pr(|\frac{R(t)}{\sqrt{Q(t)}\sigma_{B}}|>\frac{K}{\sqrt{Q(t)}\sigma_{B}}\\
 & \approx & 2\Phi(-\frac{K}{\sqrt{Q(t)}\sigma_{B}}
\end{array}
\end{equation*}

\end{IEEEproof}
We know that $\underset{t\rightarrow\infty}{\lim}Q(t)=\infty$. So
when $t\rightarrow\infty$, $R(t)$ has infinite variance, and the
over flow probability $\Pr(|R(t)|>K)\rightarrow1$. Thus, without
ENC, the system performance will decrease significantly.

This result can also be derived through Little's Law. Assume that
at the time instance $t$, the queue only has the packets from Node
A, i.e. $R(t)>0$. At the same time, packets from Node A and Node
B arrive at the relay at the identical rate of $\lambda$ packets/second.
When packets from Node A arrives, the length of queue increases. We
could see that the ``arriving rate'' is $\lambda$. When packets
from Node B arrives, the length of queue decreases with parameter
$\lambda$. So the ``service rate'' is also $\lambda$. According
to Little's Law, a queue with identical arriving time and service
time will have infinite length. Thus, with a finite buffer, the packet-loss
rate will be unavoidable.

In other case, if Node A and Node B has different flow strength, the
situation could be even worse. Assume that the strength of Node A
is $\lambda$ and $\mu$ for Node B. If $\lambda>\mu$, in the case
above, obviously the ``arriving rate'' exceeds ``service rate'',
and the queue will be infinite. If $\lambda<\mu$, when all Node A's
packets are network coded and sent out, Node B's packets start to
accumulate in the queue, which will result in $R(t)<0$, the ``arriving
rate'' exceeds ``service rate'' again. In the discussion below,
we only consider the scenario that Node A and Node B has identical
flow strength.

\begin{figure}[tb]
\centering
\includegraphics[scale=0.6]{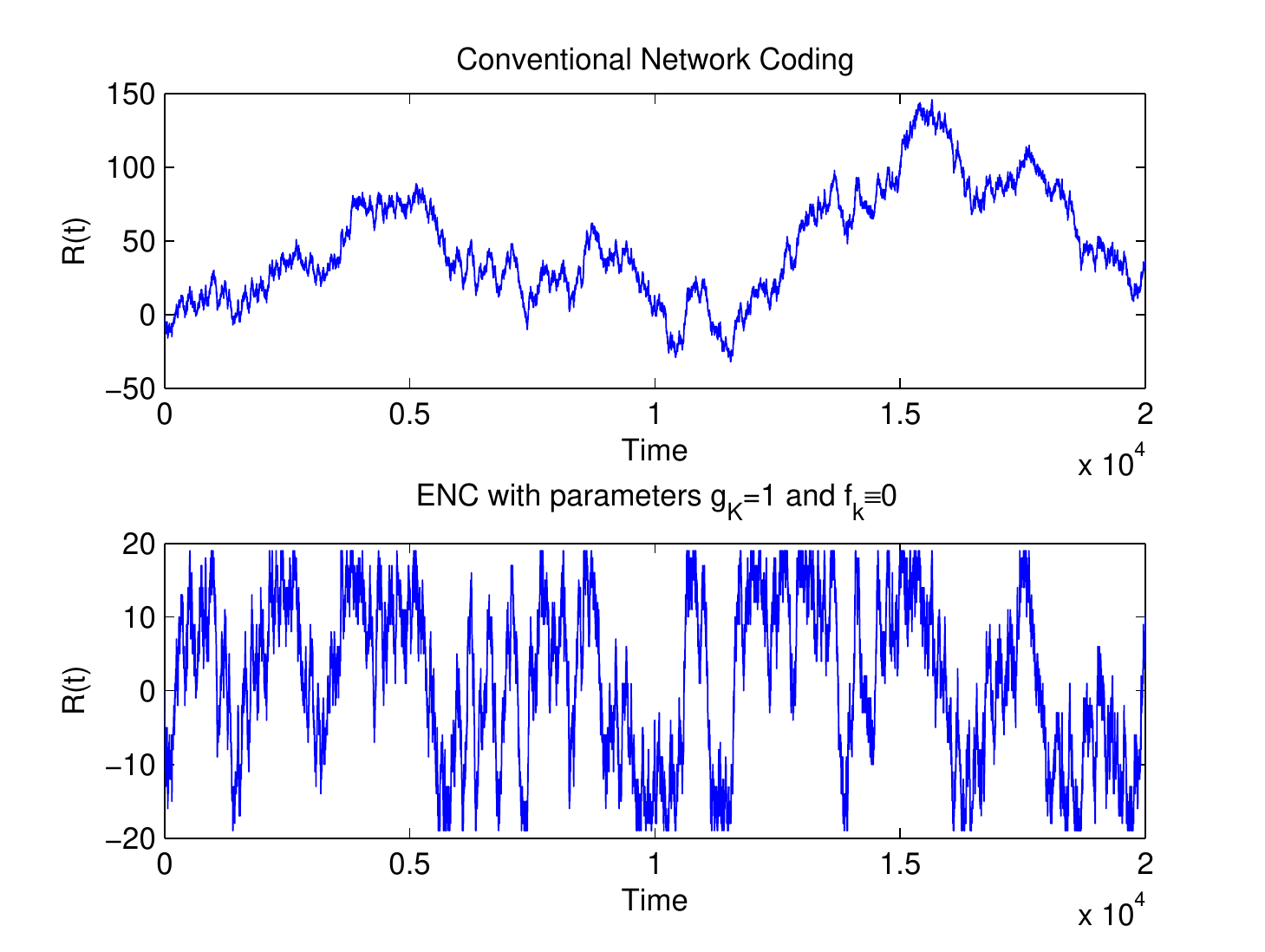}
\caption{System Performance Comparison between Conventional Network Coding
and ENC}
\label{fig:System-Performance}
\end{figure}

\autoref{fig:System-Performance} presents the system performance
comparison between conventional network coding and ENC with parameters
$g_{K}=1$ and $f_{k}\equiv0$. The buffer size $K$ is set to 20.
In this figure, we can see that in most of time, the buffer overflows
in conventional network coding. ENC limits the maximum queue length
to be 20. However, this simple group of parameters in not optimal
for ENC, since it has potential waste of energy. We will show in the
following part of this paper that there exist a group of ENC parameters
which fully utilize given energy constraint to achieve minimum delay.

\subsection{Queuing Model}

Before we considering the optimal policy on a finite buffer, we first
give a rigorous mathematical model on renewal process queuing and
its special case - Poisson process queuing.

\subsubsection{Renewal Process Queuing Model}

A renewal process is a generalization of the Poisson process. In essence,
the Poisson process is a continuous-time Markov process on the positive
integers which has independent identically distributed holding times
at each integer $i$ (exponentially distributed) before advancing
to the next integer $i+1$. In a renewal process, the distribution
of the gaps has arbitrary independent identical distribution. Let
us denote the probability distribution function as $f(t)$ and the
total number of packets as $N(t)$.
\begin{defn}
The short term average renewal rate is defined as
\end{defn}
\begin{equation*}
\lambda(t):=\frac{dE\left\{ N(t)\right\} }{dt}
\end{equation*}

\begin{defn}
\label{def:long-term-average}The long term average renewal rate is
defined as
\begin{equation*}
\lambda^{'}:=\underset{t\rightarrow\infty}{\lim}\lambda(t)
\end{equation*}
\end{defn}
With the above definitions, we notice that $\lambda^{'}$ has the
same character as the parameter $\lambda$ in a Poisson process. It
represent the number of packets arriving at the relay in a unit time.
In order to calculate $\lambda^{'}$, we have the following theorems.

\begin{lem}
\label{thm:short-term-average}
The short term average renewal rate of a renewal process is given by
\begin{equation*}
\Lambda(s)=\frac{\psi(s)}{1-\psi(s)}
\end{equation*}
where $\Lambda(s)$ and $\psi(s)$ are Laplace transformation of $\lambda(t)$
and $f(t)$ 
\begin{equation*}
\Lambda(s)=\int_{0}^{\infty}\lambda(t)\exp(-st)dt
\end{equation*}
\begin{equation*}
\psi(s)=\int_{0}^{\infty}f(t)\exp(-st)dt
\end{equation*}
\end{lem}

\begin{thm}
The long term average renewal rate of a renewal process is given by
\begin{equation*}
\lambda^{'}=\frac{1}{\int_{0}^{\infty}tf(t)dt}
\end{equation*}
\end{thm}

\begin{IEEEproof}
According to \autoref{def:long-term-average} and \autoref{thm:short-term-average},
we have
\begin{eqnarray*}
\lambda^{'} & = & \underset{t\rightarrow\infty}{\lim}\lambda(t)\\
 & = & \underset{s\rightarrow0}{\lim}s\Lambda(s)\\
 & = & \underset{s\rightarrow0}{\lim}(\frac{s}{1-\psi(s)}-s)\\
 & = & \underset{s\rightarrow0}{\lim}\frac{s}{1-\int_{0}^{\infty}f(t)\exp(-st)dt}\\
 & = & \underset{s\rightarrow0}{\lim}\frac{s}{\int_{0}^{\infty}f(t)stdt}\\
 & = & \frac{1}{\int_{0}^{\infty}tf(t)dt}
\end{eqnarray*}
Thus the above theorem holds.
\end{IEEEproof}

\subsubsection{Poisson Queuing Model}

Let us focus on the special case of renewal process - Poisson process.
Assume that the Node A and Node B exchange packets via the relay with
the rate of $\lambda$ packet/second, i.e. the short term arriving
rate $\lambda(t)$ is a constant. In the following discussion, if
we replace the Poisson parameter $\lambda$ with long term average
renewal rate $\lambda^{'}$, all theorems can be applied to renewal
process as well.

Based on the transmission policy at the relay described in 
\autoref{sec:System-Model}, the state of queues at the relay node can
be characterized with an integer $S(t)\in[0,K]$, where $K$ represents
the total length of queue at the relay.
\begin{equation*}
S(t)=|R(t)|
\end{equation*}
 Since the future state is independent from its past given the current
state, $S(t)$ is a continuous time finite state Markov chain. With
the policy defined before, we could derive the transition probability
of $S(t)$.

\begin{figure}
\begin{centering}
\includegraphics[scale=0.5]{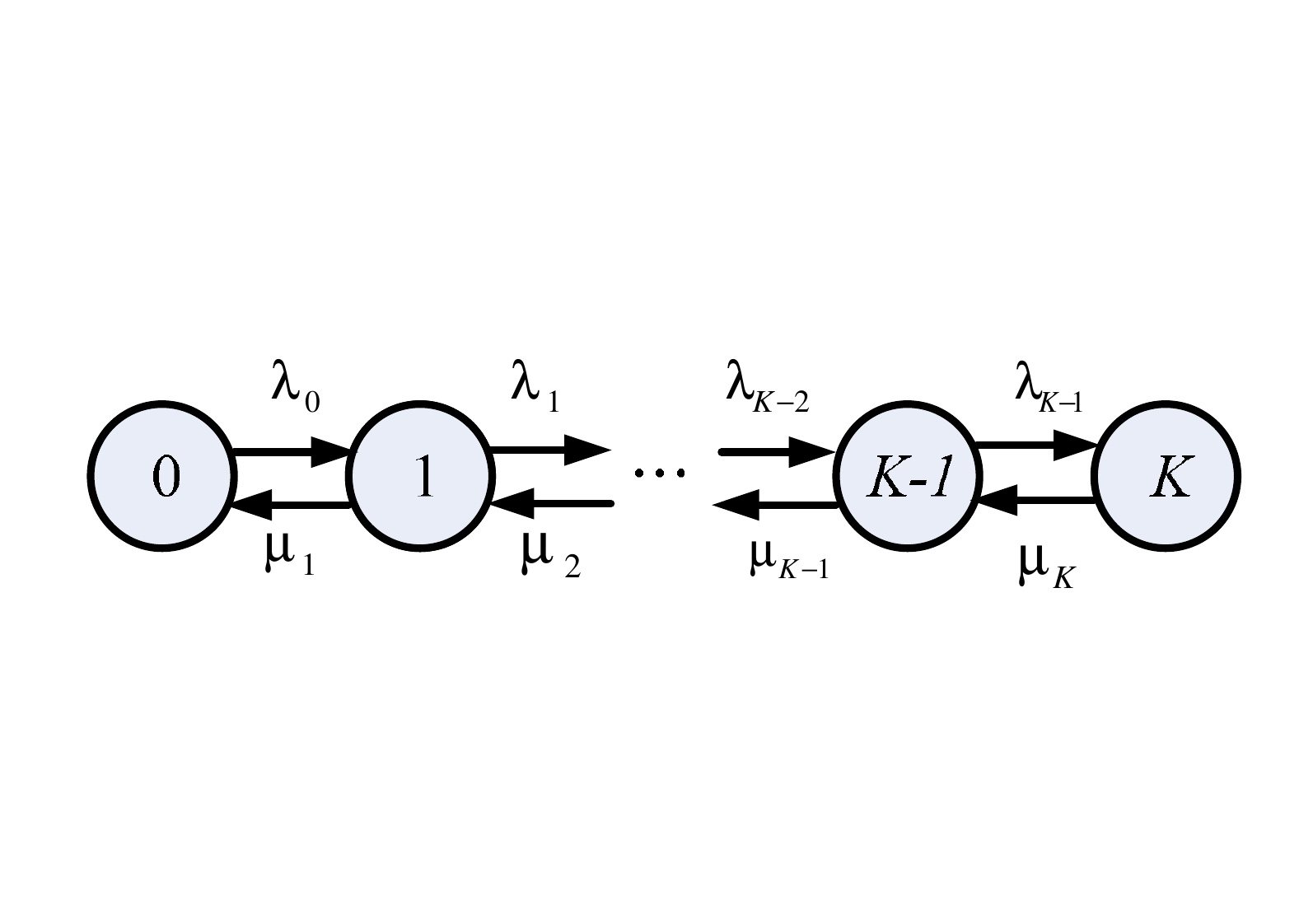}
\par\end{centering}

\begin{centering}
\caption{Finite State Markov Chain Model of Buffer State\label{fig:Finite-State-Markov}}

\par\end{centering}

\end{figure}

We depict the above mathematical description in \autoref{fig:Finite-State-Markov}.

\begin{thm}
The buffer state $S(t)$ is a $K+1$ state Markov Chain with transition
probability $\Pr(S(t+\Delta t)=j|S(t)=i)=p_{ij}\Delta t$, which satisfies
$p_{ij}=0$ for $|j-i|>1$ and
\begin{equation}
p_{k,k+1}=\begin{cases}
\lambda(1-g_{k} & 1\leq k\leq K\\
2\lambda(1-g_{0} & k=0
\end{cases}\label{eq:1}
\end{equation}

\begin{equation}
p_{k,k-1}=\lambda+f_{k}\label{eq:2}
\end{equation}
\end{thm}
\begin{IEEEproof}
First, we derive the transition probability \eqref{eq:1} and \eqref{eq:2}.
For $k=0$, the event $S(t+\Delta t)=S(t)+1$ is equivalent to the
event that the relay only receives a packet from either source node
and does not transmit an un-coded packet. Hence,
\begin{equation*}
p_{0,1}=\underset{\Delta t\rightarrow0}{\lim}\frac{2(\lambda\Delta t)(1-g_{0}}{\Delta t}=2\lambda(1-g_{0}
\end{equation*}

For $1\leq k\leq K-2$, the event $S(t+\Delta t)=S(t)+1$ is equivalent
to the event that the relay nodes only receives a packet from the
source whose packets are already in the queue, and does not transmit
an un-coded packet. Hence, we have
\begin{equation*}
p_{k,k+1}=\underset{\Delta t\rightarrow0}{\lim}\frac{(\lambda\Delta t)(1-g_{k}}{\Delta t}=\lambda(1-g_{k}
\end{equation*}

Note that the event $S(t+\Delta t)=S(t)-1$ is equivalent to the event
that (1) the relay only receives a packet from the opposite source
to one already in the queue; or (2) the relay transmits an un-coded
packet without receiving any packets. As a result,
\begin{equation*}
\begin{array}{ccc}
p_{k,k-1} & = & \underset{\Delta t\rightarrow0}{\lim}\frac{\lambda\Delta t+[1-2(\lambda\Delta t)](f_{k}\Delta t)}{\Delta t}\\
 & = & \lambda+f_{k}
\end{array}
\end{equation*}

\end{IEEEproof}
For convenience, let us denote
\begin{equation}
\lambda_{k}=\begin{cases}
p_{k,k+1} & 0\leq k\leq K-1\\
\lambda(1-g_{K} & k=K
\end{cases}\label{eq:lambdak}
\end{equation}
\begin{equation}
\mu_{k}=p_{k,k-1}\label{eq:muk}
\end{equation}

We shall mainly use $\lambda_{k}$ and $\mu_{k}$ rather than $g_{k}$
and $f_{k}$ as the ENC parameters in the following discussion. Also
note that $g_{k}\in[0,1]$ and $f_{k}\geq0$. Thus, $\lambda_{k}$
and $\mu_{k}$ satisfy the following constraints:
\begin{equation*}
0\leq\lambda_{k}\leq\begin{cases}
2\lambda & k=0\\
\lambda & 1\leq k\leq K
\end{cases}
\end{equation*}
\begin{equation*}
\mu_{k}\geq\lambda
\end{equation*}

\subsection{Delay and Packet-Loss}

With the Markov chain model, we can investigate the network layer
performance of ENC, namely, delay and packet-loss. We first present
the stationary state probability of $S(t)=k$, denoted by $\pi_{k}$,
$k=0,1,\cdots,K$, in the following theorem. The proof of this theorem
can be find in \cite{Queuing_System}.
\begin{thm}
The stationary state probability $\pi_{k}$ is given by
\begin{equation*}
\pi_{k}=\pi_{0}\overset{k-1}{\underset{m=0}{\prod}}\frac{\lambda_{m}}{\mu_{m+1}}
\end{equation*}
where
\begin{equation*}
\pi_{0}=[1+\underset{k=1}{\overset{K}{\sum}}\underset{m=0}{\overset{k-1}{\prod}}(\frac{\lambda_{m}}{\mu_{m+1}}]^{-1}\label{eq:pi0}
\end{equation*}

\end{thm}
We next present the average packet delay by the following corollary,
ignoring the potential lost packets. 
\begin{cor}
The average delay of ENC is given by
\begin{equation*}
D=\frac{1}{2\lambda}\underset{k=0}{\overset{K}{\sum}}k\pi_{k}\label{eq:4}
\end{equation*}
\end{cor}
\begin{IEEEproof}
Assume that at the time instance $t=t_{0}$, there is $k$ packets
in the queue. Let us consider the time a particular packet will experience.
Because ENC policy means no difference between two nodes, the probability
that this $k$ packets has the same source origin as our last packet
is $\frac{1}{2}$. Thus the average length of the queue consisting
of the packets from a particular node can be obtained as
\begin{equation*}
\bar{K}=\frac{1}{2}\underset{k=0}{\overset{K}{\sum}}k\pi_{k}
\end{equation*}
Recall that the average arriving time of a Poisson process is $\lambda$,
according to the Little's law, \eqref{eq:4} holds.
\end{IEEEproof}
We can also characterize the packet-loss rate of ENC.
\begin{cor}
The normalized packet-loss rate of ENC is given by
\begin{equation*}
\xi=\frac{\pi_{K}\lambda_{K}}{2\lambda}
\end{equation*}
\end{cor}
\begin{IEEEproof}
As mentioned above, the packet loss of ENC results from a full queue.
The buffer overflow event is equivalent as follows:
\begin{equation*}
\Pr(S(t+\Delta t)>K)=\Pr(S(t)=K)\lambda_{K}\Delta t
\end{equation*}
This is the buffer overflow probability for a single packet. We should
normalize it by dividing it with the total number of packets that
arrives at the relay in the time interval $\Delta t$, i.e. $2\lambda\Delta t$.
Thus, the corollary is verified.
\end{IEEEproof}
To avoid retransmission, packet dropping is not allowed in the relay
node. As a result, when the buffer is in state $S(t)=K$, the relay
must send a packet if it receives a packet from the source whose packets
are already in the queue. Thus, we have $g_{K}=1$, or equivalently,
\begin{equation*}
\lambda_{K}=0
\end{equation*}

In particular, as shown in \autoref{sub:Why-do-we}, conventional
wireless networking's packet-loss rate cannot be 0 in general. However,
in some scenario, such as UDP, packet loss is allowed. We will see
that allowing $\xi\neq0$ further reduce the energy needed.

\subsection{Energy Consumption}

Having obtained the delay and packet-loss rate of ENC, we now focus
on its energy consumption. As described in \autoref{sec:System-Model},
the relay transmits a packet under any of three circumstances. Thus,
we have the following theorem:
\begin{thm}
Conditioned on the buffer state $S(t)=k$, the energy consumption
$\varepsilon_{k}$ is a random variable satisfying
\begin{equation}
\Pr(\varepsilon_{k}=\varepsilon|S(t)=k)=\begin{cases}
-\lambda_{0}+2\lambda & k=0\\
-\lambda_{k}+\mu_{k}+\lambda & 1\leq k\leq K
\end{cases}\label{eq:5}
\end{equation}
\begin{equation}
\Pr(\varepsilon_{k}=0|S(t)=k)=1-\Pr(\varepsilon_{k}=\varepsilon|S(t)=k)\label{eq:6}
\end{equation}

where $\varepsilon$ is the average energy for transmitting one packet.\end{thm}
\begin{IEEEproof}
For $k=0$, the relay transmits with probability $g_{0}$ when receiving
packet from either source.
\begin{equation*}
\varepsilon_{0}=\varepsilon\underset{\Delta t\rightarrow0}{\lim}\frac{2(\lambda\Delta t)g_{0}}{\Delta t}=2\lambda g_{0}\varepsilon=\varepsilon(-\lambda_{0}+2\lambda)
\end{equation*}

For $1\leq k\leq K$, the relay transmits when (1) receiving a packet
from opposite source; (2) receiving a packet from the same source
whose packets are already in queue (with probability $g_{k}$); (3)
receiving no packet and transmitting the oldest packet in queue. 
\begin{equation*}
\begin{array}{ccc}
\varepsilon_{k} & =\varepsilon & \underset{\Delta t\rightarrow0}{\lim}\frac{(\lambda\Delta t)(1+g_{k}+(1-2\lambda\Delta t)(f_{k}\Delta t)}{\Delta t}\\
 & = & \varepsilon[\lambda(1+g_{k}+f_{k}]\\
 & = & \varepsilon(-\lambda_{k}+\mu_{k}+\lambda)
\end{array}
\end{equation*}
Hence, \eqref{eq:5} holds. \eqref{eq:6} can then be verified by
the probability normalization
\end{IEEEproof}
The average energy consumed in the relay node can be given by the
following theorem.
\begin{thm}
With ENC, the average energy consumed in the relay node can be given
by
\begin{equation}
E^{ave}=\varepsilon(\lambda\pi_{0}+\lambda-\lambda_{K}\pi_{K}\label{eq:result-Pave}
\end{equation}
\end{thm}
\begin{IEEEproof}
According to the definition of the average energy,
\begin{equation}
E^{ave}=\underset{k=0}{\overset{K}{\sum}}\varepsilon_{k}\pi_{k}\label{eq:definition-Pave}
\end{equation}
Then by substituting \eqref{eq:5} into \eqref{eq:definition-Pave},
it follows that
\begin{equation}
\begin{array}{ccc}
E^{ave} & = & \varepsilon[(-\lambda_{0}+2\lambda)\pi_{0}+\underset{k=1}{\overset{K}{\sum}}(-\lambda_{k}+\mu_{k}+\lambda)\pi_{k}]\\
 & = & \varepsilon[\lambda\pi_{0}+\lambda-\underset{k=0}{\overset{K-1}{\sum}}\lambda_{k}\pi_{k}+\underset{k=1}{\overset{K}{\sum}}\mu_{k}\pi_{k}-\lambda_{K}\pi_{K}]
\end{array}\label{eq:pave}
\end{equation}
According to the character of stationary probability,
\begin{equation*}
\underset{k=0}{\overset{K-1}{\sum}}\lambda_{k}\pi_{k}=\underset{k=1}{\overset{K}{\sum}}\mu_{k}\pi_{k}\label{eq:character-stationary probability}
\end{equation*}
Then by substituting \eqref{eq:character-stationary probability}
into \eqref{eq:pave}, \eqref{eq:result-Pave} holds.
\end{IEEEproof}
For different physical layer designs, the transmission energy $\varepsilon$
can be different. And for different data rate, $\lambda$ is different
as well. We notice that $E^{ave}$ is a linear function of $\varepsilon\lambda$.
To obtain a unified result, we divide $E^{ave}$ with $\varepsilon\lambda$
\begin{equation*}
\overline{E}=\frac{E^{ave}}{\varepsilon\lambda}=\pi_{0}+1-2\xi\label{eq:E-bar}
\end{equation*}

We notice that when the packet-loss rate is 0 for ENC, namely, $\xi=0$,
the normalized average energy can be further simplified to be
\begin{equation*}
\overline{E}=\pi_{0}+1
\end{equation*}

The normalized energy $\overline{E}$ represents the ratio of the
actual energy $E^{ave}$ to the average arriving rate $\lambda$ and
the average one-time energy consumption $\varepsilon$. According
to \eqref{eq:pi0}, the performance of ENC is determined by the transition
probabilities $\lambda_{k}$ and $\mu_{k}$. A key problem for ENC
is, therefore, how to choose $\lambda_{k}$ and $\mu_{k}$ to optimize
the system performance in terms of delay and energy.

\section{Optimal Delay-energy Trade-off\label{sec:Optimal-Delay-Power-Trade-off}}

We have already related the performance of ENC, in terms of average
delay and energy consumption, to the transition probabilities $\lambda_{k}$
and $\mu_{k}$, in \autoref{sec:Energy-Delay-Trade-off-Analysis}.
In this section, we shall address the fundamental problem of how can
the average delay of ENC be minimized given an average energy constraint.
The discussion is conducted in two parts - loss-free and non-loss-free.

\subsection{Loss-free Scenario}

Let $\bar{E}^{max}$ denote the normalized energy constraint. In this
context, a deterministic optimization problem for minimizing the packet
delay, $D$, of ENC can be formulated as follows.
\begin{problem}
\label{pro:Minimize}Minimize $D=\frac{1}{2\lambda}\underset{k=0}{\overset{K}{\sum}}k\pi_{k}$,
subject to

\begin{equation*}
\begin{cases}
\pi_{0}+1\leq\bar{E}^{max}\\
0\leq\lambda_{k}\leq\begin{cases}
2\lambda & k=0\\
\lambda & 1\leq k\leq K
\end{cases}\\
\mu_{k}\geq\lambda\\
\pi_{k}=\pi_{0}\overset{k-1}{\underset{m=0}{\prod}}\frac{\lambda_{m}}{\mu_{m+1}}\\
\pi_{0}=[1+\underset{k=1}{\overset{K}{\sum}}\underset{m=0}{\overset{k-1}{\prod}}(\frac{\lambda_{m}}{\mu_{m+1}}]^{-1}
\end{cases}
\end{equation*}

\end{problem}
Let $D^{*}$ denote the minimal delay determined by the above problem.
Intuitively, $D^{*}$ is a decreasing function of $\bar{E}^{max}$,which
we shall denote as
\begin{equation*}
D^{*}=d^{*}(\bar{E}^{max}
\end{equation*}

The function $d^{*}(\centerdot)$ represents the optimal delay-energy
trade-off. In this paper, we are interested in both the formulation
of the trade-off function and how to achieve the optimal trade-off.
To do this, we convert \autoref{pro:Minimize} into a linear programming
problem as described in the following theorem.
\begin{thm}
\label{thm:The-optimization-Problem}The optimization \autoref{pro:Minimize}
is equivalent to a linear programming problem given by

Minimize $D=\frac{1}{2\lambda}\pi_{0}^{*}\underset{k=1}{\overset{K}{\sum}}k\rho_{k}$,
subject to
\begin{equation*}
\begin{cases}
\underset{k=1}{\overset{K}{\sum}}\rho_{k}=\frac{1}{\pi_{0}^{*}}-1\\
\rho_{1}\leq2\\
\rho_{k+1}-\rho_{k}\leq0 & 1\leq k\leq K-1\\
\rho_{k}\geq0
\end{cases}
\end{equation*}

where $\pi_{0}^{*}$ is a constant defined by the following formula.

\begin{equation*}
\pi_{0}^{*}=\begin{cases}
0 & \bar{E}^{max}<1\\
\bar{E}^{max}-1 & 1\leq\bar{E}^{max}\leq2\\
1 & \bar{E}^{max}>1
\end{cases}
\end{equation*}
\end{thm}
\begin{IEEEproof}
Let
\begin{equation*}
\rho_{k}=\overset{k-1}{\underset{m=0}{\prod}}\frac{\lambda_{m}}{\mu_{m+1}}
\end{equation*}
It is easily seen that $\rho_{k}\geq0$. Also, we have
\begin{equation*}
\rho_{1}=\frac{\lambda_{0}}{\mu_{1}}\leq2\label{eq:rou1}
\end{equation*}
Since
\begin{equation*}
\rho_{k+1}=\frac{\lambda_{k}}{\mu_{k+1}}\rho_{k}\label{eq:rouk}
\end{equation*}
We get $\rho_{k+1}-\rho_{k}\leq0$.

Next, we note that the average delay is decreased with the increase
of the average energy. To achieve the maximum energy allowed, we could
make $\pi_{0}$ to be
\begin{equation*}
\pi_{0}=\bar{E}^{max}-1
\end{equation*}
At the same time, also note that $\pi_{0}\leq1$. Thus,
\begin{equation*}
\pi_{0}^{*}=\begin{cases}
0 & \bar{E}^{max}<1\\
\bar{E}^{max}-1 & 1\leq\bar{E}^{max}\leq2\\
1 & \bar{E}^{max}>2
\end{cases}
\end{equation*}
Then the above theorem follows.
\end{IEEEproof}
Next, we present the analytical optimal solution to \autoref{thm:The-optimization-Problem}.
\begin{thm}
The analytical optimal solution to \autoref{thm:The-optimization-Problem}
is given by
\begin{equation}
\rho_{k}^{*}=\begin{cases}
2 & k\leq k^{*}\\
\frac{1}{\pi_{0}^{*}}-1-2k^{*} & k=k^{*}+1\\
0 & k>k^{*}+1
\end{cases}\label{eq:roukstar}
\end{equation}
where
\begin{equation}
k^{*}=\left\lfloor (\frac{1}{\pi_{0}^{*}}-1)/2\right\rfloor \label{eq:kstar}
\end{equation}
\end{thm}
\begin{IEEEproof}
Note that the objective function is the weighted summation of $\rho_{k}$
with weight $k$. Subject to the given summation of $\rho_{k}$, one
needs to minimize the $\rho_{k}$ with a relatively larger weight
to minimize the weighted summation. As a result, there exists a positive
integer $k^{*}$, which satisfies
\begin{equation*}
0=\rho_{K}^{*}=\cdots=\rho_{k^{*}+2}^{*}\leq\rho_{k^{*}+1}^{*}<\rho_{k^{*}}^{*}=\cdots\rho_{1}^{*}=2
\end{equation*}
According to the constraint, $k^{*}$ is the largest integer satisfying
\begin{equation*}
2k^{*}\leq\frac{1}{\pi_{0}^{*}}-1
\end{equation*}
So the theorem holds.
\end{IEEEproof}
Having solved the linear programming problem, we next present the
optimal delay-energy trade-off function as well as the optimal ENC
that can achieve that trade-off.
\begin{thm}
The optimal ENC delay-energy trade-off function for loss-free transmission
is
\begin{equation}
\begin{array}{l}
d=\\
\begin{cases}
\frac{k^{*}+1}{2\lambda}[1-(\bar{E}^{max}-1)(k^{*}+1)] & \frac{1}{1+2K}+1\leq\bar{E}^{max}\leq2\\
0 & \bar{E}^{max}>2
\end{cases}
\end{array}\label{eq:optimal-d}
\end{equation}
\end{thm}
\begin{IEEEproof}
Notice that the maximum length of queue is $K$. We have
\begin{equation*}
\underset{k=1}{\overset{K}{\sum}}\rho_{k}\leq2K
\end{equation*}
Thus,
\begin{equation}
\bar{E}^{max}\geq1+\frac{1}{1+2K}\label{eq:E-max-constraint}
\end{equation}
Otherwise, the feasible region is empty and hence there is no solution.
Next, we substitute \eqref{eq:kstar}, \eqref{eq:roukstar} into equations
in \autoref{thm:The-optimization-Problem}, the above theorem
holds.
\end{IEEEproof}
According to \autoref{thm:The-optimization-Problem}, we can also
present the optimal parameters for ENC.
\begin{thm}
\label{thm:Final_solution}One group of optimal parameters for ENC
with maximal average energy $\bar{E}^{max}$ are given as follows:
\begin{equation}
\lambda_{k}^{*}=\begin{cases}
2\lambda & k=0\\
\lambda & 1\leq k\leq k^{*}\\
0 & k^{*}+1\leq k\leq K
\end{cases}\label{eq:lambdakstar}
\end{equation}
\begin{equation}
\mu_{k}^{*}=\begin{cases}
\lambda & k\neq k^{*}+1\\
\frac{2\lambda}{\rho_{k^{*}+1}^{*}} & k=k^{*}+1
\end{cases}\label{eq:miukstar}
\end{equation}
\end{thm}
\begin{IEEEproof}
From \eqref{eq:rou1}, \eqref{eq:rouk}, and \eqref{eq:roukstar},
we have $\lambda_{0}^{*}=2\mu_{1}^{*}$, and $\lambda_{k}^{*}=\mu_{k+1}^{*}$
for $k\leq k^{*}$. For $k>k^{*}+1$, $\rho_{k}^{*}=0$, thus $\lambda_{k-1}^{*}=0$
and $\mu_{k}^{*}$ can be any positive value. We let $\mu_{k}^{*}=\lambda$
for convenience. Finally, note that $\rho_{k^{*}}^{*}=2$, thus $\mu_{k^{*}+1}^{*}=\frac{2\lambda}{\rho_{k^{*}+1}^{*}}$.
\end{IEEEproof}
By substituting \eqref{eq:lambdakstar}, \eqref{eq:miukstar} into
\eqref{eq:lambdak}, \eqref{eq:muk}, we can obtain the optimal probability
of transmitting a packet without network coding
\begin{equation}
g_{k}^{*}=\begin{cases}
0 & k\leq k^{*}\\
1 & k^{*}+1\leq k\leq K
\end{cases}\label{eq:gkstar}
\end{equation}
\begin{equation}
f_{k}^{*}=\begin{cases}
\frac{2\lambda}{\rho_{k^{*}+1}^{*}}-\lambda & k=k^{*}+1\\
0 & k\neq k^{*}+1
\end{cases}\label{eq:fkstar}
\end{equation}

This is the analytical result of delay-energy trade-off curve lower
bound. ENC on this parameters could achieve the delay lower bound
given energy constraint.

\subsection{Non-loss-free Scenario}

\begin{figure}[tb]
\centering
\includegraphics[scale=0.6]{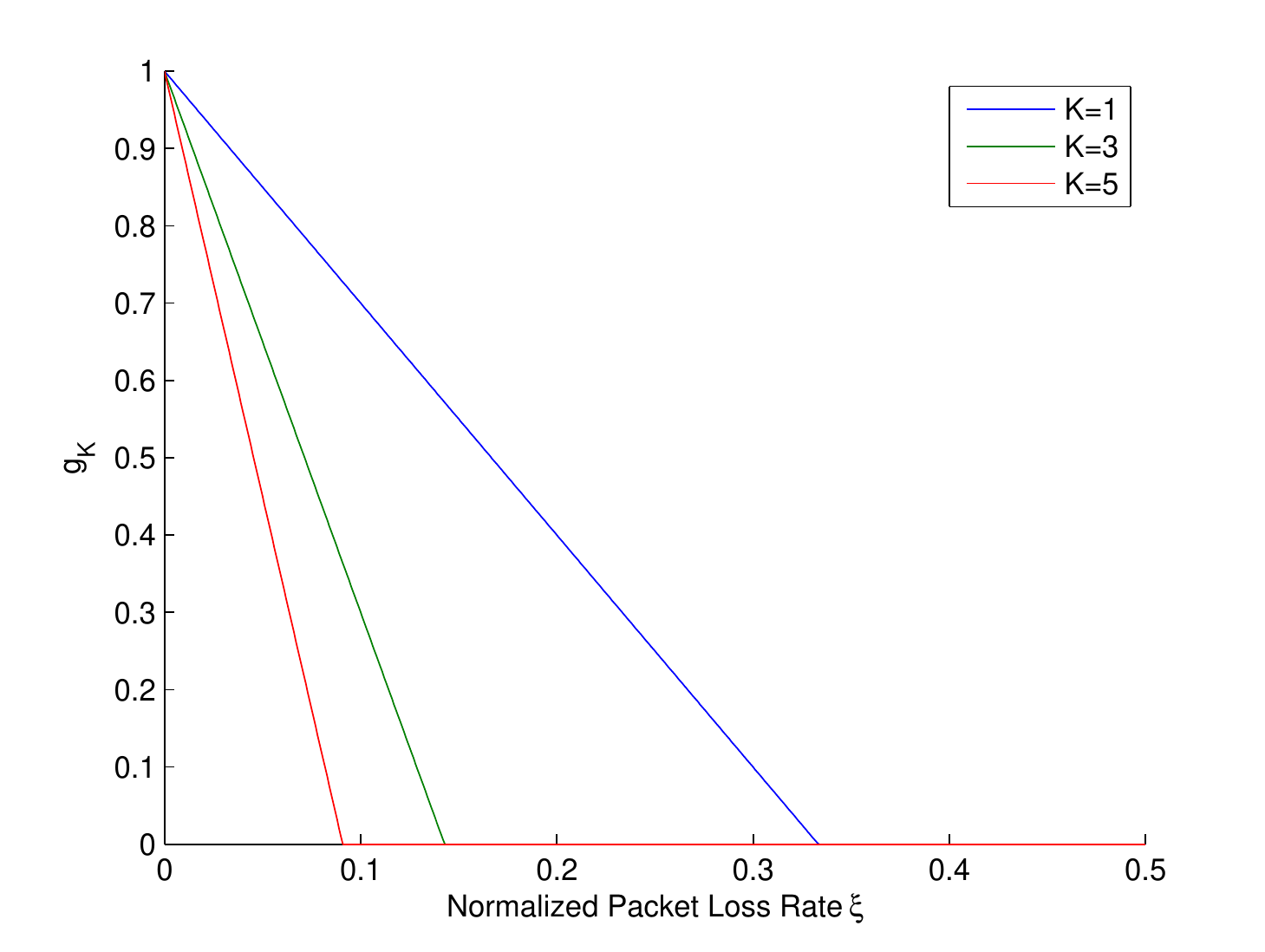}
\caption{Packet-loss Rate of ENC}
\label{fig:Packet-loss-Rate}
\end{figure}

In this subsection, the optimal delay function and ENC parameters
that satisfy the allowed normalized packet loss rate $\xi$ are presented.
We assume that the packet-loss in ENC only is only introduced by a
full queue. Note that the delay in non-loss-free scenario only represents
the delay of the packets that arrive at the destination successfully.
\begin{thm}
The optimal delay-energy trade-off function that achieve normalized
packet loss rate $\xi$ in a $K$-packet buffer is
\begin{equation}
d=\frac{K}{2\lambda}[1-(\bar{E}^{max}-1+2\xi)K]\label{eq:optimal-d-loss}
\end{equation}
where $\frac{1}{1+2K}+1-2\xi\leq\bar{E}^{max}\leq\frac{1}{1+2K}+1$.\end{thm}
\begin{IEEEproof}
Recall that in \eqref{eq:E-max-constraint}, we require that $\bar{E}^{max}\geq1+\frac{1}{1+2K}$.
However, this constraint is not always satisfied. \eqref{eq:E-bar}
implies that the normalized packet loss rate $\xi$ can be view as
a virtual energy that lowers down the energy threshold from $1+\frac{1}{1+2K}$
to $1+\frac{1}{1+2K}-2\xi$. Moreover, in this case, from \eqref{eq:roukstar},
$\rho_{K}=\frac{1}{\pi_{0}}-1-2(K-1)=2$. Thus, $k^{*}+1=K$ according
to \eqref{eq:kstar}. The optimal delay function can be applied to
non-loss-free scenario simply by replacing $\bar{E}^{max}$ with $\bar{E}^{max}+2\xi$
in \eqref{eq:optimal-d} and replacing $k^{*}+1$ with $K$.
\end{IEEEproof}
In the non-loss-free scenario, we are not interested system performance
in terms of delay and energy. We are more interested in how to configure
ENC's parameters - especially $g_{k}$ to achieve particular packet
loss rate.
\begin{thm}
The parameters of ENC that achieve normalized packet loss rate $\xi$
in a $K$-packet buffer with energy constraint $\bar{E}^{max}=1+\frac{1}{1+2K}-2\xi$
is
\begin{equation*}
g_{k}=\begin{cases}
1 & 0\leq k\leq K-1\\
1-(1+2K)\xi & k=K
\end{cases}\label{eq:gk_conventional}
\end{equation*}
\begin{equation*}
f_{k}\equiv0
\end{equation*}
\end{thm}
\begin{IEEEproof}
According to \eqref{eq:E-bar},

\begin{equation*}
\pi_{0}=\bar{E}^{max}-1+\xi=\frac{1}{1+2K}
\end{equation*}
As we already know in the proof to the last theorem, $\rho_{K}=2$,
we have $\pi_{K}=\pi_{0}\rho_{K}=\frac{2}{1+2K}$. Recall the definition
of normalized packet loss rate,
\begin{equation*}
\pi_{K}\lambda_{K}=2\xi\lambda\label{eq:normalized-packet-loss-rate}
\end{equation*}
By substitute \eqref{eq:normalized-packet-loss-rate} into \eqref{eq:lambdak},
the theorem holds.
\end{IEEEproof}
\autoref{fig:Packet-loss-Rate} shows the relationship between
normalized packet loss rate $\xi$ and ENC parameter $g_{K}$. We
could see that $g_{K}$ changes linearly against $\xi$. However,
one can not save energy in exchange of packet loss rate unlimitedly.
It is because the stationary probability of a full queue $\pi_{K}$
is not large enough for introducing a packet loss rate of $\xi$.
When $\xi$ exceeds $\frac{1}{1+2K}$, i.e. the normalized energy
constraint is less than $1-\frac{1}{1+2K}$, ENC can not achieve the
minimal energy constraint by enhancing allowed packet loss rate constraint. 

\section{Numerical Results}
\label{sec:Numerical-Results}

In this section, sample numerical results are presented to demonstrate
the potential of ENC and validate the theoretical results of this
work. We developed a event-driven simulator with C++ based on Monte-Carlo
method to verify our theoretical model.

\subsection{Packet Loss Rate for Conventional Network Coding}

\begin{figure}[tb]
\centering
\includegraphics[scale=0.6]{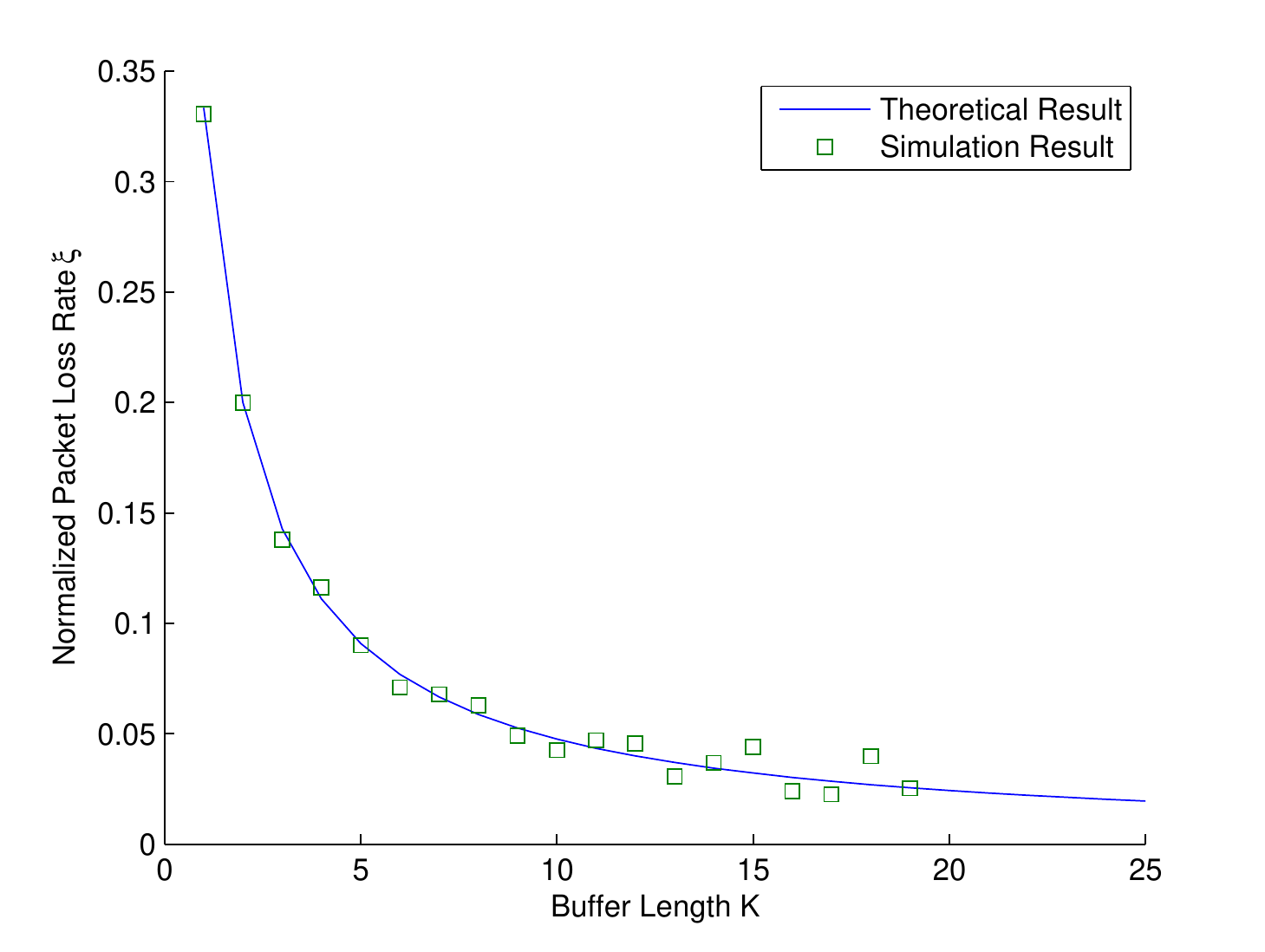}
\caption{Packet Loss Rate for Conventional Network Coding}
\label{fig:Conventional-Packet-Loss}
\end{figure}

In \eqref{eq:gk_conventional}, if we let $g_{K}=0$, the ENC reduces
to conventional network coding. We are interested in acquiring the
curve of normalized packet-loss rate $\xi$ against the buffer length
$K$ when $g_{K}=0$. \autoref{fig:Conventional-Packet-Loss} presents
the theoretical curve of $\xi=\frac{1}{1+2K}$ and simulation result
in squares. 

Again, we observe that with a finite buffer, conventional network
coding will result in inevitable packet loss, i.e. $\xi\neq0$. However,
when conventional network coding scheme employs a larger buffer, the
packet loss rate will decreases.

\subsection{Delay-Energy Trade-off for ENC}

Assume that the maximal relay buffer space is $K=3$. The theoretical
optimal delay-energy trade-off curve, as well as the simulation results
of the optimal trade-off achieving ENC strategies, is presented. 

\begin{figure}[h]
\begin{centering}
\includegraphics[scale=0.6]{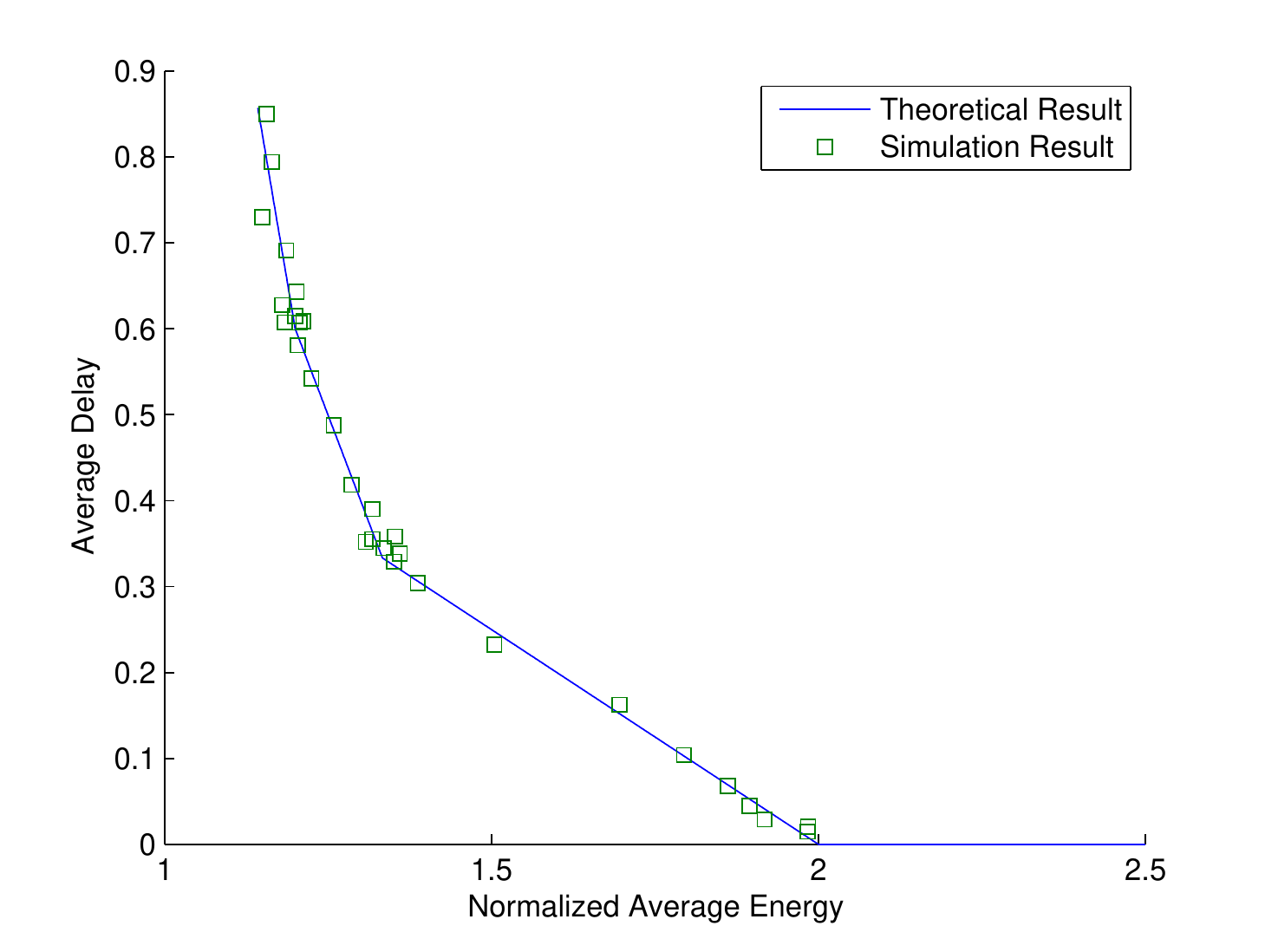}
\par\end{centering}

\noindent \centering{}\caption{Optimal Delay-energy Trade-off for ENC\label{fig:Optimal-Delay-Power-Trade-off}}
\end{figure}

We observe the delay and energy consumed while changing $f_{k}^{*}$
and $k^{*}$. \autoref{fig:Optimal-Delay-Power-Trade-off} shows
the theoretical and simulation results for the optimal delay-energy
trade-off. The theoretical optimal trade-off curve is shown by the
solid line. The simulation results for the four optimal ENC strategies
are show by the squares in the three cases. It can be seen that the
simulation results fit on the theoretical curve perfectly.

From this result, we could see that
\begin{enumerate}
\item The optimal delay is a decreasing function of minimal energy constraint.
The curve consists $K$ straight lines, which indicates the delay
is linear to the minimal energy constraint.
\item In the simulation, notice the three special cases, where the energy
constraints are given by
\begin{equation*}
\begin{array}{cc}
\bar{E}_{m}^{max}=1+\frac{1}{1+2m}+\delta & m=1,2,3\end{array}
\end{equation*}
where $\delta\rightarrow0^{+}$. Recall that in \eqref{eq:roukstar},
\eqref{eq:gkstar} and \eqref{eq:fkstar}, 
\begin{equation*}
\rho_{k^{*}+1}^{*}=\frac{1}{\pi_{0}^{*}}-1-2\left\lfloor (\frac{1}{\pi_{0}^{*}}-1)/2\right\rfloor \in[0,2]
\end{equation*}
When $\bar{E}_{m}^{max}$ is \emph{slightly larger than} $1+\frac{1}{1+2m}$,
$\rho_{k^{*}+1}^{*}=2$, $f_{k^{*}+1}^{*}=0$. So we have $k^{*}+1=m$
and $f_{k}^{*}\equiv0$ for any $k$ in this case. The optimal ENC
strategy becomes quite simple. 
\item When the energy constraint is large, namely, $\bar{E}^{max}>2$, the
delay is zero. That means the relay transmits a packet the time receiving
it without storing. When the energy constraint is small, the delay
increases rapidly. When $\bar{E}^{max}<1+\frac{1}{1+2K}=1+\frac{1}{7}\approx1.143$,
the delay is infinite, which indicates the given energy is not sufficient
for packet-loss free transmitting. Recall that in \autoref{sub:Why-do-we},
we prove that conventional network coding could not avoid packet loss
in a finite buffer. Here we can easily see that the conventional network
coding has the normalized energy constraint $\bar{E}^{max}=1<1+\frac{1}{1+2K}$
for any positive integer $K$.
\end{enumerate}

\section{Conclusions}

In this paper, we proposed a continuous-time opportunistic network
coding or ENC method in the continuous time domain, where the relay
node can transmit either network-coded or un-coded packets. We show
that the conventional network coding scheme on a finite buffer implies
inevitable packet loss. By using a Markov model, the packet delay,
packet-loss rate, and average energy consumption of ENC were presented
for general renewal process queuing and classical Poisson process
queuing. Given the delay and energy results, the performance bound
of ENC was proposed in terms of delay-energy trade-off. We also presented
an ENC strategy which can achieve the optimal trade-off. Through simulations,
the developed theoretical results were validated. It was also demonstrated
that the proposed ENC can achieve lower delay compared to conventional
wireless network coding scheme.

\bibliographystyle{IEEEtran}
\bibliography{paper}

\end{document}